\pdfoutput=1
\documentclass[11pt]{article}

\usepackage[final]{acl}

\usepackage{times}
\usepackage{latexsym}
\usepackage{hyperref}
\usepackage{microtype}
\usepackage{graphicx}
\usepackage{subfigure}
\usepackage{booktabs} % for professional tables

\usepackage{listings}
\usepackage{xcolor}
\usepackage{float}
\usepackage{verbatim}

\usepackage{tcolorbox} 
\usepackage[inline]{enumitem}
\setlist[itemize]{leftmargin=4mm}
\setlist[enumerate]{leftmargin=4mm}
\newfloat{listing}{H}{lop}

\usepackage[T1]{fontenc}
\usepackage[utf8]{inputenc}

\usepackage{microtype}

\usepackage{inconsolata}

\usepackage{graphicx}

\title{\as{}: A No-Code Developer Tool for Building and Debugging Multi-Agent Systems}

\author{\\ Victor Dibia,
  Jingya Chen,
  Gagan Bansal,
  Suff Syed, \\
  Adam Fourney,
  Erkang Zhu,
  Chi Wang,
  Saleema Amershi \\
  Microsoft Research, Redmond, United States \\
  \texttt{\{victordibia, jingyachen, gaganbansal, suffsyed, adam.fourney,  } \\
  \texttt{erkang.zhu, chiw, samershi\}@microsoft.com}
}

\begin{document} 
\newcommand{\lida}{\textsc{lida}} 
\newcommand{\igm}{\textsc{igm}} 
\newcommand{\nl}{\textsc{nl}} 
\newcommand{\llm}{\textsc{llm}} 
\newcommand{\ma}{\textsc{multi-agent}} 
\newcommand{\vizbench}{\textsc{VizBench}}
\newcommand{\autogena}{\textsc{AutoGen Assistant}}
\newcommand{\autogen}{\textsc{AutoGen}}
\newcommand{\as}{\textsc{AutoGen Studio}}
\newcommand{\ags}{\textsc{AutoGen Studio}}
\newcommand{\agsdownloads}{\textsc{200k}}
\newcommand{\agsissues}{\textsc{135}}

\definecolor{myOrange}{rgb}{1,0.5,0}
\definecolor{myPurple}{HTML}{4045DB}
\definecolor{myLightPurple}{HTML}{D3C3F6}
\definecolor{myLightGray}{HTML}{F0F0F0}

\definecolor{myPurple}{HTML}{4045DB}\definecolor
{myLightYellow}{HTML}{FADA62}

\newcommand{\gbox}[3]{ 
   
    \begin{tcolorbox}[width=.999\columnwidth,  colback={#3}, halign=left] 
        \textbf{#1} 
        \tcblower
        {#2}
      \end{tcolorbox}
}

\newtcbox{\mcbox}[1][red]{on line,
arc=5pt,colback=#1!50!white,colframe=#1!90!black,
before upper={\rule[-3pt]{0pt}{10pt}},boxrule=1pt,
boxsep=0pt,left=6pt,right=6pt,top=2pt,bottom=2pt}

\newenvironment{metaverbatim}{\verbatim}{\endverbatim} 

\newcommand{\gboxx}[2]{ 
    \begin{tcolorbox}[width=.999\columnwidth,  colback={#2}, halign=left] 
        {#1}
    \end{tcolorbox}
}

\maketitle

\begin{abstract}
Multi-agent systems, where multiple agents (generative AI models + tools) collaborate, are emerging as an effective pattern for solving long-running, complex tasks in numerous domains. However, specifying their parameters (such as models, tools, and orchestration mechanisms etc,.) and debugging them remains challenging for most developers. 
To address this challenge, we present \as{}, a no-code developer tool for rapidly prototyping, debugging, and evaluating multi-agent workflows built upon the \autogen{} framework. \as{} offers a web interface and a Python API for representing \llm{}-enabled agents using a declarative (JSON-based) specification. It provides an intuitive drag-and-drop UI for agent workflow specification, interactive evaluation and debugging of workflows, and a gallery of reusable agent components. We highlight four design principles for no-code multi-agent developer tools and contribute an open-source implementation.\footnote{\url{https://github.com/microsoft/autogen/tree/autogenstudio/samples/apps/autogen-studio}}
 
\end{abstract}
\section{Introduction}

When combined with the ability to act (e.g., using tools),  Generative AI models function as agents, enabling complex problem-solving capabilities. Importantly, recent research has shown that transitioning from prescribed (fixed) agent pipelines to a multi-agent setup with autonomous capabilities can result in desirable behaviors such as improved factuality and reasoning \cite{du2023improving}, as well as divergent thinking \cite{liang2023encouraging}. These observations have driven the development of application frameworks such as AutoGen \cite{wu2023autogen}, CAMEL \cite{li2024camel}, and TaskWeaver \cite{qiao2023taskweaver}, which simplify the process of crafting multi-agent applications expressed as Python code. However, while multi-agent applications advance our capacity to solve complex problems, they also introduce new challenges. For example, developers must now configure a large number of parameters for these systems including defining agents (e.g., the model to use, prompts, tools or skills available to the agent, number of action steps an agent can take, task termination conditions etc.), communication and orchestration mechanisms - i.e., the order or sequence in which agents act as they collaborate on a task. Additionally, developers need to debug and make sense of complex agent interactions to extract signals for system improvement. All of these factors can create significant barriers to entry and make the multi-agent design process tedious and error-prone.
To address these challenges, we have developed \as{}, a tool for rapidly prototyping, debugging, and evaluating \ma{} workflows. Our contributions are highlighted as follows:

\begin{figure}[t]
    \centering
    \includegraphics[width=\columnwidth]{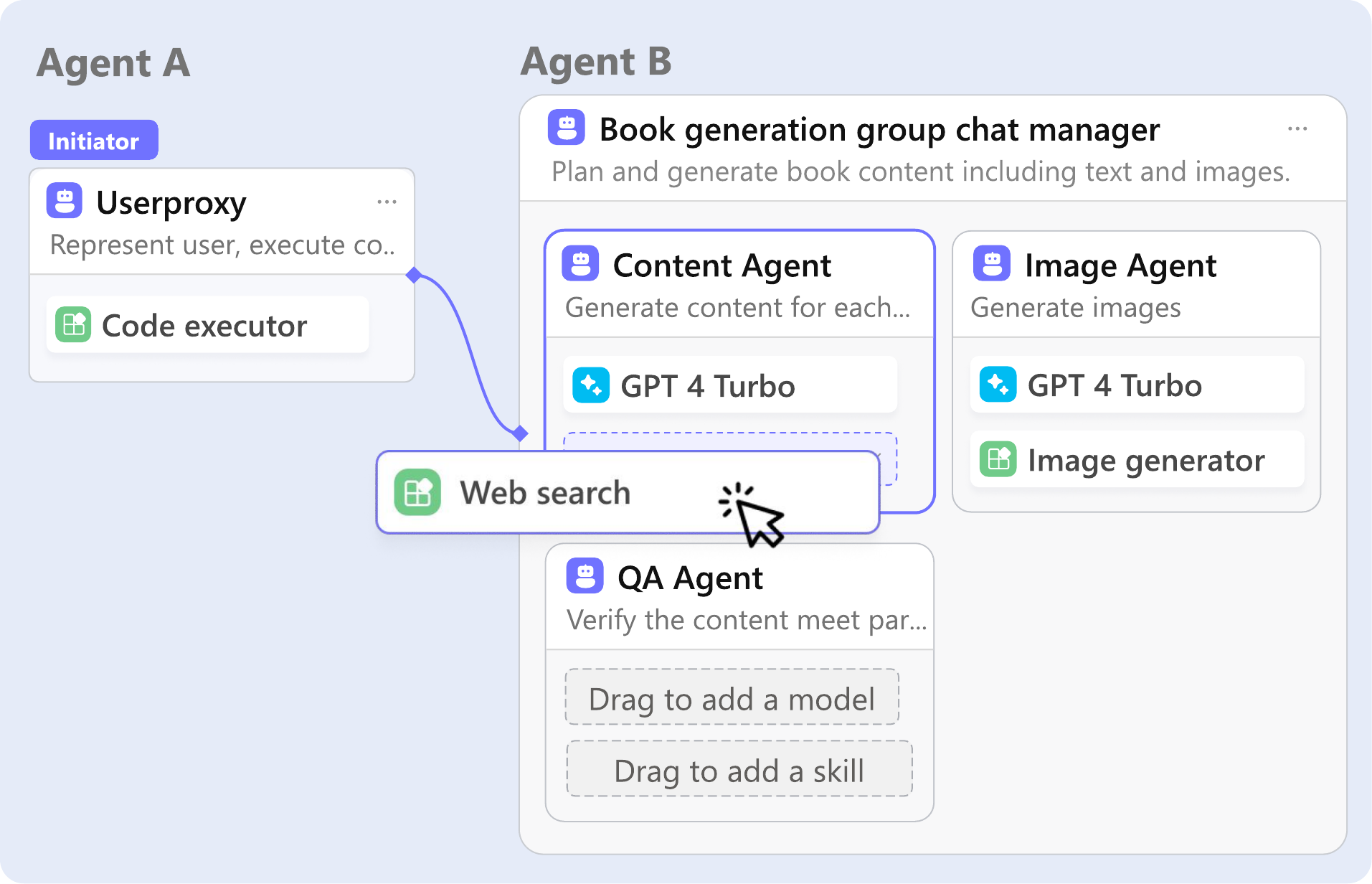}
    \caption{\as{} provides a drag-n-drop UI where models, skills/tools, memory components can be defined,  \textit{attached} to agents and agents \textit{attached} to workflows.}
    \label{fig:ags_dragdrop}
\end{figure}

\begin{itemize}
    \item \as{} - a developer-focused tool (UI and backend Web and Python API) for declaratively specifying and debugging (human-in-the-loop and non-interactive) \ma{} workflows. \as{} provides a novel  drag-and-drop experience (Figure \ref{fig:ags_dragdrop}) for rapidly authoring complex \ma{} agent workflows, tools for profiling/debugging agent sessions, and a gallery of reusable/shareable \ma{} components.
    \item We introduce profiling capabilities with visualizations of messages/actions by agents and metrics (costs, tool invocations, and tool output status) for debugging \ma{} workflows.
    \item Based on our experience building and supporting \as{} as an open-source tool with a significant user base (over \href{https://www.pepy.tech/projects/autogenstudio}{\agsdownloads{} downloads}   within a 5-month period), we outline emerging design patterns for \ma{} developer tooling and future research directions.
\end{itemize}

To the best of our knowledge, \ags{} is the first open-source project to explore a no-code interface for autonomous \ma{} application development, providing a suitable platform for research and practice in \ma{} developer tooling. 
\section{Related Work}

\subsection{Agents ( \llm{s} + Tools)}
Generative AI models face limitations, including hallucination — generating content not grounded in fact — and limited performance on reasoning tasks or novel out-of-distribution problems. To address these issues, practice has shifted towards agentic implementations where models are given access to tools to act and augment their performance \cite{mialon2023augmented}.  Agentic implementations, such as React \cite{yao2022react}, explore a Reason and Act paradigm that uses LLMs to generate both reasoning traces and task-specific actions in an interleaved manner. As part of this process, developers have explored frameworks that build prescriptive pipelines interleaving models and tools (e.g., LIDA \cite{dibia2023lida}, LangChain \cite{Chase_LangChain_2022}). However, as tasks become more complex, requiring lengthy context  and the ability to independently adapt to dynamic problem spaces, predefined pipelines demonstrate limited performance \cite{liu2024lost}. This limitation has led to the exploration of more flexible and adaptive agent architectures.

\subsection{\ma{} Frameworks}
Several frameworks have been proposed to provide abstractions for creating such applications. AutoGen \cite{wu2023autogen} is an open-source extensible framework that allows developers to build large \ma{} applications. CAMEL \cite{li2024camel} is designed to facilitate autonomous cooperation among communicative agents through role-playing, using inception prompting to guide chat agents toward task completion while aligning with human intentions. OS-Copilot \cite{wu2024copilot} introduces a framework for building generalist agents capable of interfacing with comprehensive elements in an operating system, including the web, code terminals, files, multimedia, and various third-party applications. It explores the use of a dedicated planner module, a configurator, and an executor, as well as the concept of tools ( Python functions or calls to API endpoints) or skills (tools that can be learned and reused on the fly). 
% Voyager \cite{wang2305voyager} is an embodied lifelong learning agent in Minecraft that continuously explores the world, acquires diverse skills, and makes novel discoveries without human intervention. It consists of three key components: an automatic curriculum that maximizes exploration, an ever-growing skill library for storing and retrieving complex behaviors, and a new iterative prompting mechanism that incorporates environment feedback, execution errors, and self-verification for program improvement. 
% TaskWeaver \cite{qiao2023taskweaver} is a code-first agent framework that seamlessly plans and executes data analytics tasks by converting user requests into executable code and treating user-defined plugins as callable functions. 

\gbox{Multi-Agent Core Concepts}{
  \label{definition:multiagentconcepts}
  \begin{enumerate}
    \item \textbf{Model}: Generative AI model used to drive core agent behaviors. 
    \item \textbf{Skills/Tools}: Code or APIs used to address specific tasks. 
    \item \textbf{Memory}: Short term (e.g., lists) or long term (vector databases) used for to save and recall information.
    \item \textbf{Agent}: A configuration that ties together the model, skills, memory components and behaviors.
    \item \textbf{Workflow}: A configuration of a set of agents and how they interact to address tasks (e.g., order or 
sequence in which agents act, task planning, termination conditions etc.). 
  \end{enumerate} 
}{myLightGray}

Collectively, these tools support a set of core capabilities  - definition of \textit{\textbf{agent}} parameters -  such as generative AI \textit{\textbf{models}}, \textit{\textbf{skills / tools}} or memory, and agent \textit{\textbf{workflows}} - specifications of how these agents can collaborate. However, most of these frameworks primarily support a code-first representation of agent workflows, which presents a high barrier to entry and rapid prototyping. They also do not provide tools or metrics for agent debugging and evaluation. Additionally, they lack structured reusable templates to bootstrap or accelerate the agent workflow creation process. \as{} addresses these limitations by providing a visual interface to declaratively define and visualize agent workflows, test and evaluate these workflows, and offer templates for common \ma{} tasks to streamline development. While this work is built on the \autogen{} open source library \cite{wu2023autogen} and inherits the core abstractions for representing agents, the proposed design patterns on no-code developer tools are intended to apply to all \ma{} frameworks.

\section{Design Goals}
\as{} is designed to enhance the \ma{} developer experience by focusing on three core objectives:

\noindent \textbf{Rapid Prototyping:} Provide a \textit{playground} where developers can quickly specify agent configurations and compose these agents into effective multi-agent workflows.
    
\noindent \textbf{Developer Tooling:} Offer \textit{tools} designed to help developers understand and debug agent behaviors, facilitating the improvement of multi-agent systems.
    
\noindent \textbf{Reusable Templates:} Present a gallery of reusable, shareable templates to bootstrap agent workflow creation. This approach aims to establish shared standards and best practices for \ma{} system development, promoting wider adoption and implementation of \ma{} solutions.

% These objectives collectively aim to streamline the development process, enhance system understanding, and foster a collaborative ecosystem for \ma{} applications.
\section{System Design}
\label{sec:systemdesign}

\as{} is implemented across two high-level components: a frontend user interface (UI) and a backend API (web, python and command line). It can be installed via the PyPI package manager (listing \ref{listing:ui_install}).

\begin{figure*}[t!]
    \centering
    \includegraphics[width=\textwidth]{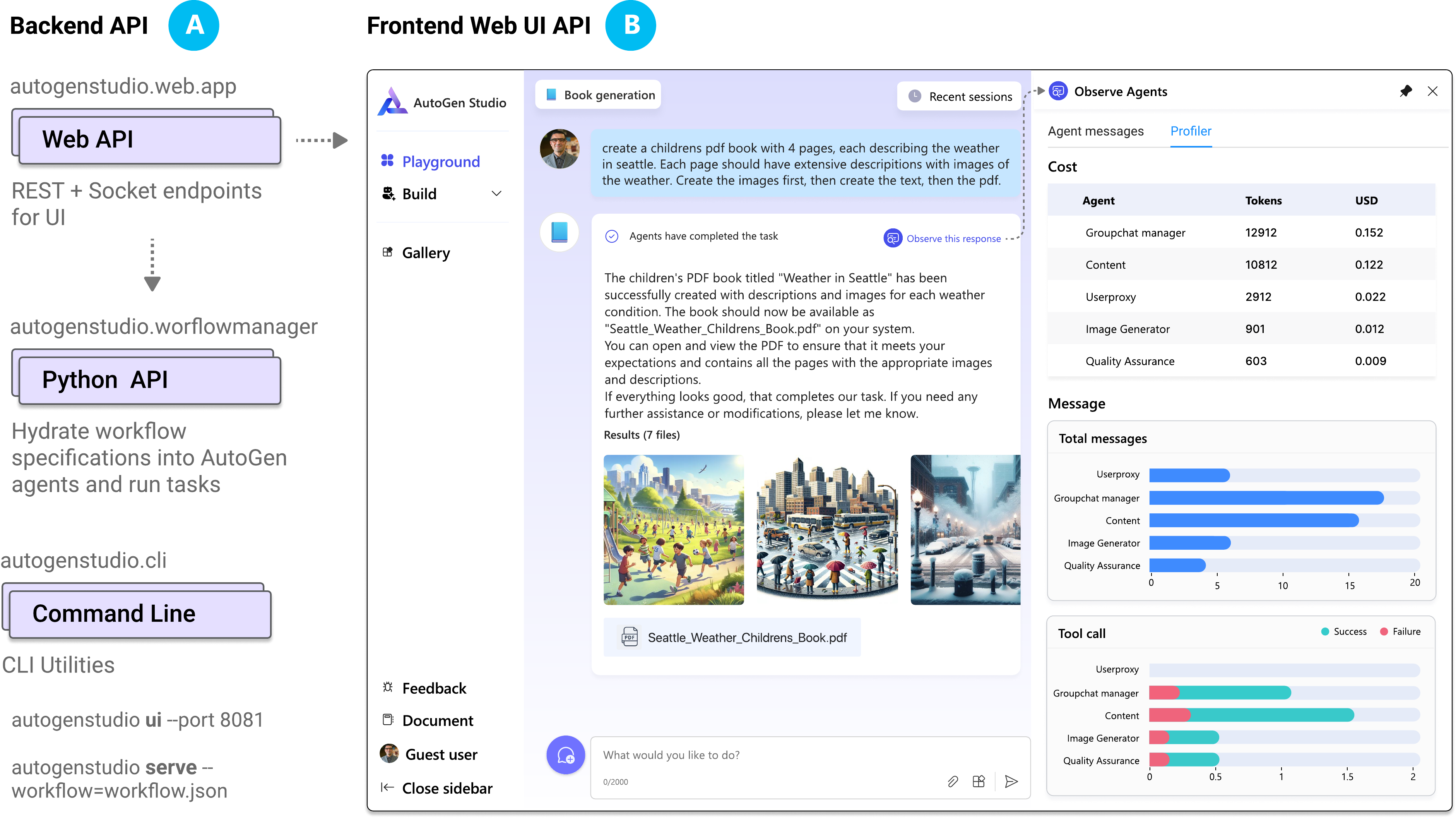}
    \caption{ \as{} provides a backend api (web, python, cli) and a UI which implements a  \textit{playground} (shown), \textit{build} and \textit{gallery} view. In the playground view, users can run tasks in a session based on a workflow. Users can also \textit{observe} actions taken by agents, reviewing agent messages and metrics based on a profiler module. }
    \label{fig:ags_architecture}
\end{figure*}

\begin{listing}[H]
\begin{tcolorbox}[width=\columnwidth, colback={myLightGray}, halign=left,left=2pt,
  right=0pt,
  top=0pt,
  bottom=0pt,]
\begin{lstlisting}[language=bash,
  numbers=none,
  frame=none]
pip install autogenstudio
autogenstudio ui --port 8081 
\end{lstlisting}
\end{tcolorbox}
\caption{\ags{} can be installed from PyPI (pip) and the UI launched from the command line.}
\label{listing:ui_install}
\end{listing}

\subsection{User Interface}
The frontend web interface in \ags{} is built using React and implements three main views that support several key functionalities. The \textit{build view} enables users to author (define-and-compose) multi-agent workflows. The \textit{playground view} allows for interactive task execution and workflow debugging, with options to export and deploy. The \textit{gallery view} facilitates the reuse and sharing of agent artifact templates.

\subsubsection{Building Workflows}
The \textit{build} view in the UI (see Figure \ref{fig:ags_dragdrop}) offers a \textit{define-and-compose} experience, allowing developers to declaratively define low-level components and iteratively compose them into a workflow. For instance, users can define configurations for models, skills/tools (represented as Python functions addressing specific tasks), or memory stores (e.g., documents organized in a vector database). Each entity is saved in a database for use across interface interactions. Subsequently, they can define an agent, attaching models, skills, and memory to it. Several agent default templates are provided following \autogen{} abstractions - a \textit{UserProxy} agent (has a code execution tool by default), an \textit{AssistantAgent} (has a generative AI model default), and a \textit{GroupChat} agent (an abstraction container for defining a list of agents, and how they interact). Finally, workflows can be defined, with existing agents attached to these workflows. The default workflow patterns supported are autonomous chat (agents exchange messages and actions across conversation turns until a termination condition is met) and sequential chat (a sequence of agents defined, each agent processes its input in order and passes on a summary of their output to the next agent). The workflow composition process is further enhanced by supporting a drag-and-drop interaction e.g., skills/models can be dragged to agents and agents into workflows.

\subsubsection{Testing and Debugging Workflows}

Workflows can be tested in-situ in the \textit{build} view, or more systematically explored within the \textit{playground} view. The playground view allows users create \textit{sessions}, attach workflows to the session, and run tasks (single shot or multi-turn). Sessions can be shared (to illustrate workflow performance) and multiple sessions can be compared. \ags{} provides two features to support debugging. First, it provides an observe view\textit{} where as tasks progress, messages and actions performed by agents are streamed to the interface, and all generated artifacts are displayed (e.g., files such as images, code, documents etc). Second a post-hoc profiler view is provided where a set of metrics are visualized for each task addressed by a workflow - total number of messages exchanged, costs (generative AI model $tokens$ consumed and dollar costs), how often agents use tools and the $status$ of tool use (success or failure), for each agent.

\subsubsection{Deploying Workflows }

\ags{} enables users to export workflows as a JSON configuration file. An exported workflow can be seamlessly integrated into any Python application (listing \ref{listing:workflow_python}), executed as an API endpoint using the \ags{} command line interface (figure \ref{fig:ags_architecture}a), or wrapped in a Docker container for large-scale deployment on various platforms (Azure, GCP, Amazon, etc.).

\begin{listing}[H]
\begin{tcolorbox}[width=\columnwidth, colback={myLightGray}, halign=left,left=2pt,
  right=0pt,
  top=0pt,
  bottom=0pt,]
\begin{lstlisting}[language=Python,
  numbers=none,
  frame=none]
from autogenstudio import WorkflowManager
wm = WorkflowManager("workflow.json")
wm.run(message="What is the height of the Eiffel Tower")
\end{lstlisting}
\end{tcolorbox}
\caption{Workflows can be imported in python apps.}
\label{listing:workflow_python}
\end{listing}

% \begin{listing}[H]
% \begin{tcolorbox}[width=\columnwidth, colback={myLightGray}, halign=left,left=2pt,
%   right=0pt,
%   top=0pt,
%   bottom=0pt,]
% \begin{lstlisting}[language=bash,
%   numbers=none,
%   frame=none]
% autogenstudio serve --workflow=workflow.json
% \end{lstlisting}
% \end{tcolorbox}
% \caption{Workflows can be run as API endpoints.}
% \label{listing:workflow_cli}
% \end{listing}

% \begin{tcolorbox}[width=.999\columnwidth,  colback={myLightGray}, halign=left,, title={Serve workflow as web api endpoint}] 
% \begin{metaverbatim}
% autogenstudio serve workflow.json 
% \end{metaverbatim}
% \label{listing:workflow_commandline}
% \end{tcolorbox}

% \begin{metaverbatim}
% autogenstudio serve --workflow=workflow.json
%  --port=5000
% \end{metaverbatim}

% \begin{listing}[H]
% \begin{lstlisting}[language=Python, caption={Python code to load and run a workflow}]
% from autogenstudio import WorkflowManager
% workflow_manager = WorkflowManager(workflow="path/to/your/workflow_.json")
% workflow_manager.run(message="What is the height of the Eiffel Tower")
% \end{lstlisting}
% \end{listing}

% The workflow can be launched as an API endpoint from the command line using the \texttt{autogenstudio} command-line tool.

% \begin{listing}[H]
% \begin{lstlisting}[language=bash, caption={Command to serve the workflow as an API endpoint}]
% autogenstudio serve --workflow=workflow.json --port=5000
% \end{lstlisting}
% \end{listing}

% Finally, the workflow launch command above can be wrapped into a Dockerfile that can be deployed on cloud services like Azure Container Apps or Azure Web Apps.

\subsubsection{Template Gallery}

The UI also features a \textit{gallery} view - a repository of components (skills, models, agents, workflows) that users can import, extend, and reuse in their own workflows. Since each component specification is declarative (JSON), users can also easily export, version and reshare them.
% \paragraph{Important Note:}
% Currently, skills, models, agents, and workflows are stored as independent, unlinked entities within a database. When an entity is incorporated into another (e.g., adding a skill to an agent), a snapshot of the entity is created and linked to the target entity. Subsequent changes to the original entity do not propagate to the linked entity, necessitating manual updates.

\subsection{Backend API - Web, Python, and Command Line}
The backend API comprises three main components: a web API, a Python API, and a command-line interface. The web API consists of REST endpoints built using the FastAPI library\footnote{FastAPI: https://fastapi.tiangolo.com/}, supporting HTTP GET, POST, and DELETE methods. These endpoints interact with several key classes:
A $DBManager$ performs CRUD (Create, Read, Update, Delete) operations on various entities such as skills, models, agents, memory, workflows, and sessions. The $WorkflowManager$ class handles the ingestion of declarative agent workflows, converts them into \autogen{} agent objects, and executes tasks (see listing \ref{listing:workflow_python}). A $Profiler$ class parses agent messages to compute metrics. When a user initiates a task within a session, the system retrieves the session history, instantiates agents based on their serialized representations from the database, executes the task, streams intermediate messages to the UI via websocket, and returns the final results.
\ags{} also provides a command-line interface with utilities for launching the bundled UI and running exported workflows as API endpoints.

\section{Usage and Evaluation}

\begin{figure}[t]
    \centering
    \includegraphics[width=\columnwidth]{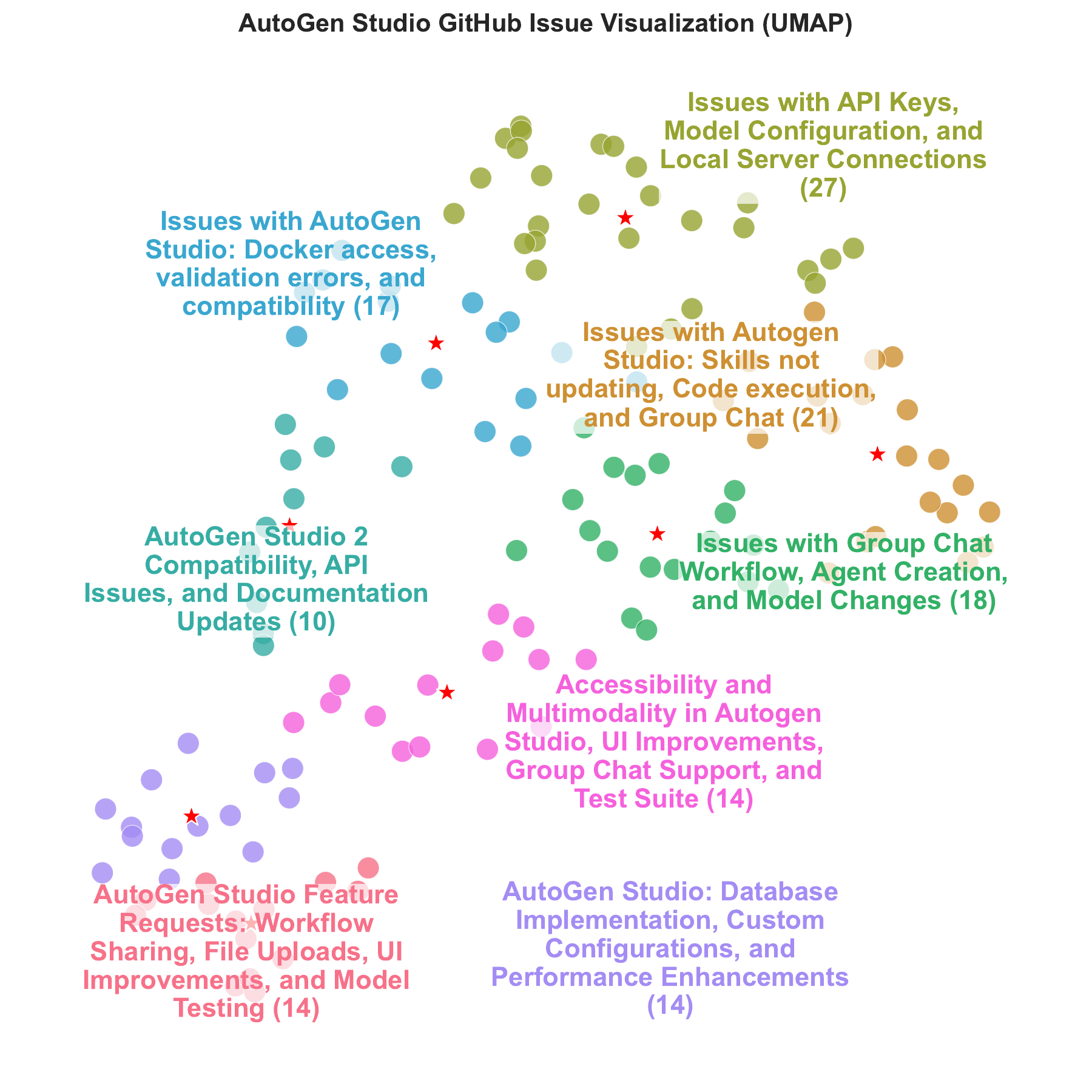}
    \caption{Plot of GitHub issues ($n=8$ clusters) from the \ags{} repo. User feedback ranged from support with workflow authoring tools (e.g., the ability configure and test models)  to general installation. }
    \label{fig:ags_issueplot}
\end{figure}

In this project, we have adopted an in-situ, iterative evaluation approach. Since its release on GitHub (5 months), the \as{} package has been installed over \href{https://www.pepy.tech/projects/autogenstudio}{\agsdownloads{} times} and has been iteratively improved based on feedback from usage (> \agsissues{} GitHub \href{https://github.com/microsoft/autogen/issues?q=is%3Aissue+label%3Astudio}{issues}). Issues highlighted several user pain points that were subsequently addressed including: (a) challenges in defining, persisting, and reusing components, resolved by implementing a database layer; (b) difficulties in authoring components, resolved by supporting automated tool generation from descriptions and integrating an IDE for editing tools; (c) frustrations caused by components failing during end-to-end tests, addressed by incorporating a test button for components (e.g.,models) and workflows in the \textit{build} view. Figure \ref{fig:ags_issueplot} displays a plot of all \ags{} issues. Each point represents an issue, based on an embedding of its text (title + body) using OpenAI's \href{https://platform.openai.com/docs/guides/embeddings}{text-embedding-3-large} model. The embeddings were reduced to two dimensions using UMAP, clustered with K-Means ($k=8$), and cluster labels generated using GPT-4 (grounded on 10 samples from its centroid). Finally, in Appendix \ref{sec:appendixb}, we demonstrate how \ags{} can effectively be used to support an engineer persona in rapidly prototyping, testing, and iteratively debugging a \ma{} workflow, and deploying it as an API endpoint to address a concrete task (generating books).

\section{Emerging Design Patterns and Research Directions}

 In the following section, we outline some of the high-level emerging patterns which we hope can help inform the design of no-code interfaces for building next-generation multi-agent applications.

\subsection{Define-and-Compose Workflows}

\gboxx{
  \label{p:1}
   Allow users to author workflows by defining components and composing them (via drag-and-drop actions) into multi-agent workflows.
}{myLightGray}

A multi-agent system can have a wide array of parameters to configure. We have found that selecting the right visual presentation of the workflow to helping users understand what parameters to configure (discovery), and how to configure them. Specifically, we have found that a define-and-compose workflow, where entities are first defined and persisted independently, and then composed ultimately into multi-agent workflows, provides a good developer experience. This includes providing tools to support authoring entities e.g., the ability define and test models, an IDE for generating/editing tools (code), and a a canvas-based visual layout of workflows with drag-and-drop interaction for associating entities in the workflow.

\subsection{Debugging and Sensemaking Tools} 
\gboxx{
  \label{p:2}
  Provide robust tools to help users debug, interpret, and rationalize the behavior and outputs of multi-agent systems.
}{myLightGray}

Multi-agent workflows can be brittle and fail for multiple reasons, ranging from improperly configured models to poor instructions for agents, improper tool configuration for agents or termination conditions. A critical request has been for tools to help users debug and make sense of agent responses.

\subsection{Export and Deployment}
\gboxx{
  \label{p:3}
  Enable seamless export and deployment of multi-agent workflows to various platforms and environments.
}{myLightGray}
While a no-code tool like \ags{} enables rapid iteration and demonstration of workflows, the natural progression for most use cases is that developers want to replicate the same outcomes but integrated as parts of their core applications. This stage requires seamless export and deployment of multi-agent workflows to various platforms and environments.

\subsection{Collaboration and Sharing}
\gboxx{
  \label{p:4}
  Facilitate user collaboration on multi-agent workflow development and allow easy sharing of creations within the community.
}{myLightGray}

Collaboration and sharing are key to accelerating innovation and improving multi-agent systems. By enabling users to collaborate on workflow development, share their creations, and build upon each other’s work, a more dynamic and innovative development environment can be cultivated. Tools and features that support real-time collaboration, version control, and seamless sharing of workflows and components are essential to foster a community-driven approach. Additionally, offering a repository or gallery where users can publish and share their workflows, skills, and agents promotes communal learning and innovation.

\section{Future Research Directions}

While we have explored early implementations of the design requirements mentioned above, our efforts in building \ags{} have also identified two important future research areas and associated research questions.

\begin{itemize}
    \item \textbf{Offline Evaluation Tools}: This encompasses questions such as how can we measure the performance, reliability, and reusability of agents across tasks? How can we better understand their strengths and limitations? How can we explore alternative scenarios and outcomes? And how can we compare different agent architectures and collaboration protocols?
    
    \item \textbf{Understanding and quantifying the impact of multi-agent system design decisions}: These questions include determining the optimal number and composition of agents for a given problem, the best way to distribute responsibilities and coordinate actions among agents, and the trade-offs between centralized and decentralized control or between homogeneous and heterogeneous agents. 
    
    \item \textbf{Optimizing of multi-agent systems}: Research directions here include the dynamic generation of agents based on task requirements and available resources, tuning workflow configurations to achieve the best performance, and adapting agent teams to changing environments and user preferences. Furthermore, how can we leverage human oversight and feedback to improve agent reliability, task performance and safety?
\end{itemize}

\section{Conclusion}
This paper introduced \as{}, a no-code developer tool for rapidly prototyping, debugging, and evaluating multi-agent workflows. Key features include a drag-and-drop interface for agent workflow composition, interactive debugging capabilities, and a gallery of reusable agent components. Through widespread adoption, we identified emerging design patterns for multi-agent developer tooling - a define and compose approach to authoring workflows, debugging tools to make sense of agent behaviors, tools to enable deployment and collaborative sharing features.
\as{} lowers the barrier to entry for multi-agent application development, potentially accelerating innovation in the field. Finally we outline future research directions including developing offline evaluation tools, ablation studies to quantify the impact of \ma{} systems design decisions and methods for optimizing multi-agent systems.
\section{Ethics Statement}

\ags{} is designed to provide a no-code environment for rapidly prototyping and testing multi-agent workflows. Our goal is to responsibly advance research and practice in solving problems with multiple agents and to develop tools that contribute to human well-being. Along with \autogen{}, \ags{} is committed to implementing features that promote safe and reliable outcomes. For example, \ags{} offers profiling tools to make sense of agent actions and safeguards, such as support for Docker environments for code execution. This feature helps ensure that agents operate within controlled and secure environments, reducing the risk of unintended or harmful actions. For more information on our approach to responsible AI in AutoGen,  please refer to transparency FAQS  \href{https://github.com/microsoft/autogen/blob/main/TRANSPARENCY_FAQS.md}{here}. Finally, \ags{} is not production ready i.e., it does not focus on implementing authentication and other security measures that are required for production ready deployments.

\section*{Acknowledgements}

We would like to thank members of the open-source software (OSS) community and the AI Frontiers organization at Microsoft Research for discussions and feedback along the way. Specifically, we would like to thank Piali Choudhury, Ahmed Awadallah, Robin Moeur, Jack Gerrits, Robert Barber, Grace Proebsting, Michel Pahud, Qingyun Wu, Harsha Nori and others for feedback and comments.

\bibliography{paper}

\appendix
\newpage
% \onecolumn
% \section{System Implementation}
% \label{sec:appendixa}

% This is a section in the appendix.

% \begin{figure*}[ht!]
%     \centering
%     \includegraphics[width=\textwidth]{figures/ags_system_arch.pdf}
%     \caption{ The \as{} architecture is composed of 2 high level components (a) Frontend UI that allows the user to rapidly specify (build view), evaluate and debug multi-agent workflows (playground sessions view), reuse and share artifacts (gallery view) (b) A Web API that supports the underlyin UI actions, a Python API (WorkflowManager) consumed by the web api that interacts with the  \autogen{} library to enable agent behaviors.}
%     \label{fig:ags_build}
% \end{figure*}

\section{Jack the Software Engineer Persona Use Case}

% \begin{figure*}[!t]
%     \centering
%     \includegraphics[width=\textwidth]{figures/Jack_Userflow.pdf}
%     \caption{Jack's user flow of creating a multi-agent system for Book Generation Application.}
%     \label{fig:ags_build}
% \end{figure*}

% \begin{figure*}[!t]
%     \centering
%     \includegraphics[width=\textwidth]{figures/ags_export_gal.png}
%     \caption{\textit{Gallery view}: user could utilize reusable examples or publish their works in the gallery.}
%     \label{fig:ags_gallery}
% \end{figure*}

\label{sec:appendixb}
Jack is a junior software engineer who has recently joined SoftwareCon. As part of his tasks, he is required to create an application that can generate a variety of short books. The initial version should focus on generating children's books (age 5 -8 years old) based on a given query (e.g., \textit{create a book for kids on how the sun works}) with the expectation of being generalized to support other generic tasks. Jack has heard about a \ma{} approach to building systems that can address a variety of tasks through autonomous collaboration between agents. To explore this approach, he begins by perusing the \ags{}  documentation, installs it, launches the UI, and performs the following steps:

\subsection{Step 1: Define and Compose a Workflow}
 
Jack starts with the \textit{Build} view, where he reviews the default skills that come with \ags{}. He sees that there are two relevant skills $generate\_pdfs$ and $generate\_images$. He verifies that he has the appropriate API keys for the $generate\_image$ skill. Next, he creates a GPT3.5 model and adds an API key.
    
Following best practices, Jack knows that the basic agent team with \autogen{} consists of a \textit{UserProxyAgent} that can execute code and an \textit{AssistantAgent} that can solve tasks as well as write code or call available tools/skills. He creates both of these agents; for his \textit{AssistantAgent}, he ensures that he attaches the GPT4 model he created previously and also attaches both skills. Jack moves on to the workflow tab and creates a new autonomous chat workflow where he specifies the \textit{UserProxyAgent} as the initiator and his \textit{AssistantAgent} as the receiver. 
    
\subsection{Step 2: Test and Iterate}
Within the workflow tab, Jack tests the workflow immediately and quickly observes a few issues. Using the profiler tool and visualization of messages exchanged by the agents, he notices that there seem to be quality issues with the content of the book - namely, the \textit{AssistantAgent} seems to generate very short messages and hence the book pages contains only 2 sentences per page whereas the requirements state that the kids are slightly older and can read much longer text.
    
To remedy these issues, Jack takes two actions. First, he attempts to extend the base instructions of his \textit{AssistantAgent}, but still doesn't get pages with more than 3 sentences across interactive tests. He recalls that using more agents can help separate focus and improve task performance. He then switches to creating 4 agents: a \textit{UserProxy}, a \textit{ContentAssistant} with detailed instructions on generating the content for each page, a \textit{QualityAssuranceAssistant} to verify the pages meet parameters, and an \textit{ImageGeneratorAssistant} focused on generating images for the book. He then creates a \textit{GroupChat} agent and adds his list of agents to it. Next, he creates a new workflow where the receiver is the \textit{GroupChat} agent and tests the application across a few tries. Jack is satisfied with the results as full-page stories are now generated correctly. In addition, Jack is concerned about costs but can easily use the \textit{observe message} button to explore duration, tokens used by agents, tool/skill use and LLM dollar costs for each task run.
    
\subsection{Step 3: Export and Share}
At this point, Jack has two final tasks: he wants to share his work with colleagues for feedback and then provide an API they can prototype with. \ags{} makes sharing easy;  First, Jack can simply export and share a link to successful sessions. Second, he can also download his workflow and share it with colleagues, saving it in a version control system like Git. Third, he can spin up an API endpoint where the agents can respond to task requests using cli commands `autogenstudio serve --port 8000`. He can also spin up a docker container using the \ags{} serve command and scale it on any platform of his choice (Azure, AWS, GCP, Hugging Face).

\end{document}